\documentclass[aps,prl,reprint,groupedaddress]{revtex4-1}

\usepackage{graphicx}

\begin{document}


\title{Geometric Phase and Non-Adiabatic Effects in an Electronic Harmonic Oscillator}

\author{M. Pechal}
\email[]{mpechal@phys.ethz.ch}
\author{S. Berger}
\author{A.A. Abdumalikov Jr.}
\author{J.M. Fink}
\author{J.A. Mlynek}
\author{L. Steffen}
\author{A. Wallraff}
\author{S. Filipp}
\affiliation{Department of Physics, ETH Zurich, CH-8093 Zurich, Switzerland}

\date{\today}

\begin{abstract}

Steering a quantum harmonic oscillator state along cyclic trajectories leads to a path-dependent geometric phase. Here we describe an experiment observing this geometric phase in an electronic harmonic oscillator. We use a superconducting qubit as a non-linear probe of the phase, otherwise unobservable due to the linearity of the oscillator.
Our results demonstrate that the geometric phase is, for a variety of cyclic trajectories, proportional to the area enclosed in the quadrature plane. At the transition to the non-adiabatic regime, we study corrections to the phase and dephasing of the qubit caused by qubit-resonator entanglement. The demonstrated controllability makes our system a versatile tool to study adiabatic and non-adiabatic geometric phases in open quantum systems and to investigate the potential of geometric gates for quantum information processing.

\end{abstract}


\maketitle

It is a~well known result in quantum mechanics that an adiabatically changing Hamiltonian causes a~system, prepared initially in a~stationary state, to follow the instantaneous energy eigenstates. This \emph{adiabatic theorem} \cite{MessiahII} determines the evolution of the state up to a~complex phase factor. Starting in an energy eigenstate, a~cyclic adiabatic change of the Hamiltonian drives the system along a~closed path in the space of physically distinct states. As noted by Berry \cite{berry1984}, even when corrected for the familiar dynamical phase given by the energy of the system, the initial and final state after the cyclic evolution can differ by a~geometric phase determined only by the path. Geometric effects have been experimentally observed in a variety of two-level systems such as single neutrons \cite{bitter1987}, nuclear spins \cite{suter1988,jones2000} and superconducting qubits \cite{leek2007,Neeley2009}. The geometric phase of another simple quantum system - a harmonic oscillator - has been used for entangling gates using harmonic motional modes of trapped ions \cite{leibfried2003} but has not been a subject of detailed experimental investigations in other systems. 

The independence of the geometric phase on dynamical quantities makes it stable under certain types of fluctuations in the system parameters \cite{dechiara2003,carollo2004,whitney2005}, offering interesting possibilities for potential noise-resilient quantum gates. Although the properties of geometric phases in the presence of noise in more general settings are still under debate \cite{blais2003,thomas2011}, geometric gates have been proposed in various physical implementations \cite{fuentesguridi2000, wang2001A, zheng2004, garcia-ripoll2005, blais2007,zanardi1999, duan2001, pachos2002,kamleitner2011} and noise-induced geometric dephasing has been studied experimentally \cite{leek2007,filipp2009G,cucchietti2010}.

Here we describe an experiment measuring the geometric phase of an adiabatically manipulated harmonic oscillator in an electronic superconducting circuit \cite{blais2004}. In contrast to anharmonic systems such as qubits, the linearity of the harmonic oscillator implies that the accumulated geometric phase is the same for all energy eigenstates and therefore cannot be measured simply by observing the phase difference between two adiabatically transported eigenstates in a~superposition \cite{leek2007}. Instead, we utilize a dispersive coupling between a qubit and the oscillator which introduces a~shift of the resonator frequency depending on the qubit state. The difference between the geometric phases accumulated for the two oscillator frequencies is then measured using the qubit as an interferometer, as also proposed in \cite{Vacanti2011}. The good controllability of our system allows us to investigate the phase for a~wide range of parameters. In particular, we identify corrections to the phase and dephasing of the qubit caused by qubit-resonator entanglement at the transition to non-adiabatic evolution. In this regime, fast geometrically protected gates based on the Aharonov-Anandan phase \cite{Aharonov1987}, which are not restricted by the adiabaticity condition, may be realizable \cite{zhuPRL2002}.

In our setup, the harmonic oscillator is implemented as one of the electromagnetic modes of a~transmission line resonator at a~frequency of $\omega_r/2\pi \approx 7.0\ \mathrm{GHz}$ (with the qubit in the ground state). It is dispersively coupled to a~superconducting qubit of the transmon type \cite{koch2007} with an energy separation between the two lowest lying energy levels of $\hbar\omega_q \approx h\times 8.3\ \mathrm{GHz}$, an anharmonicity $\alpha/2\pi\approx -0.4\ \mathrm{GHz}$ and a resonant coupling strength to the oscillator of $g/2\pi\approx 56\ \mathrm{MHz}$. The sample is operated in a~dilution refrigerator at a~base temperature of approximately $10\ \mathrm{mK}$.

We manipulate the state of the resonator using a~microwave drive field. The Hamiltonian in the reference frame rotating at the drive frequency $\omega$ is
\[
  H = \hbar (\omega_r - \omega) a^{\dagger} a + 
  \frac{1}{2}\hbar\varepsilon_I(t) (a^{\dagger} + a) +
  \frac{1}{2}\hbar i\varepsilon_Q(t) (a^{\dagger} - a),
\]
where $\varepsilon_I(t)$ and $\varepsilon_Q(t)$ are the in-phase and quadrature components of the drive, which we can control individually. These two components are directly related to the amplitude $\varepsilon$ and phase $\varphi$ of the drive by $\varepsilon_I + i\varepsilon_Q = \varepsilon e^{i\varphi}$.

\begin{figure}
\includegraphics{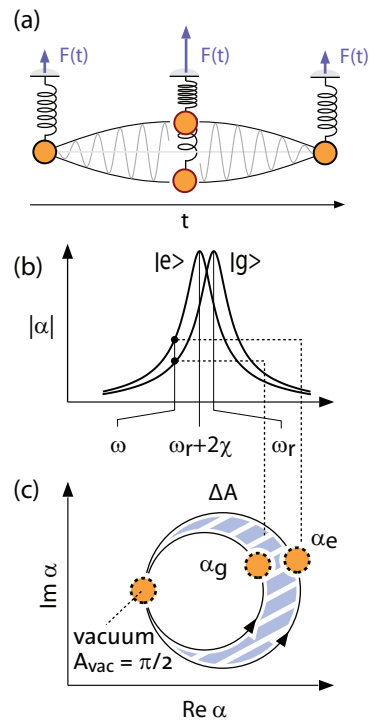}
\caption{
{(a)}
Adiabatically driven mechanical oscillator formed by a mass on a~spring. Displacement is proportional to the quickly oscillating force with slowly varying amplitude and phase.
{(b)}
The Lorentzian resonator response function centered at frequencies $\omega_r$ and $\omega_r+2\chi$ for the qubit in $|g\rangle$ and $|e\rangle$, respectively.
{(c)}
Area of the coherent state path in the complex quadrature plane depending on the qubit state. The measured geometric phase difference is proportional to the area $\Delta A$ between the paths. Dashed circles with an area $A_{\mathrm{vac}}=\pi/2$ represent the size (rms) of vacuum fluctuations.
}
\label{fig1}
\end{figure}

If the variation of $\varepsilon_I(t)$ and $\varepsilon_Q(t)$ is slow compared with the detuning $\delta = \omega_r - \omega$, the resonator field adiabatically follows the ground state of the Hamiltonian. This situation is analogous to the oscillations of a mechanical harmonic oscillator, such as a mass on a spring, following the changes in amplitude and phase of a driving force [Fig.~\ref{fig1}(a)]. The ground state is a~coherent state $|\alpha\rangle$, that is, an eigenstate of the annihilation operator $a$ with eigenvalue $\alpha = -(\varepsilon_I + i\varepsilon_Q) / 2\delta$. Classically, the absolute value and phase of $\alpha$ correspond to the amplitude and phase of the resonator field, respectively. Its square $|\alpha|^2$ is the mean number of photons $n$ in the resonator.

We can manipulate the coherent state of the resonator at will by changing the drive components $\varepsilon_I$ and $\varepsilon_Q$. These are calibrated in terms of the resonator photon number using a~qubit ac~Stark shift measurement \cite{schuster2007}. If we make its quadrature $\alpha$ trace a~closed path in the complex plane, returning back to the vacuum state $\alpha = 0$, the system acquires a~dynamical phase proportional to the time integral of $\varepsilon_I^2+\varepsilon_Q^2$ and a~geometric phase $-2A$, where $A$ is the area enclosed by the path \cite{chaturvedi1987}. 

The field quadrature $\alpha$ depends on the detuning $\delta$ between the drive and the resonator frequency through the Lorentzian response function of the resonator characterized by its center frequency $\omega_r$ and its width $\kappa/2\pi\approx 500\ \mathrm{kHz}$ [Fig.~1(b)]. The dressed resonator frequency is determined by the state of the dispersively coupled qubit \cite{blais2004}. We denote it by $\omega_r$ for the qubit in its ground state $|g\rangle$. If the qubit is in the excited state $|e\rangle$, the resonator frequency is shifted to $\omega_r+2\chi$. The size of the path traced by the coherent state, and hence also the accumulated phase $\gamma^{(s)}$, therefore depends on the qubit state $|s\rangle$ [Fig.~\ref{fig1}(c)].

\begin{figure}
\includegraphics{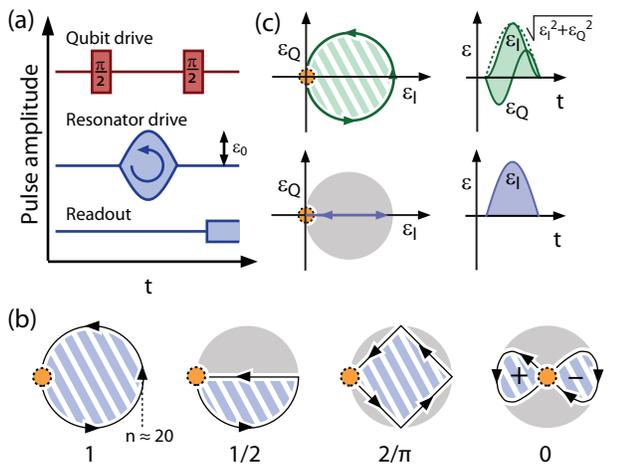}
\caption{
{(a)}
Illustration of the qubit and resonator drive pulses. The maximum amplitude of the resonator drive is denoted by $\varepsilon_0$.
{(b)}
The four different shapes of the path used in the experiment and the corresponding geometric phases relative to the one of the circular path.
{(c)}
Path of the resonator drive components in the $IQ$ plane (left) and the corresponding time dependence (right) of the components $\varepsilon_I$ and $\varepsilon_Q$ (solid line) and the drive amplitude $(\varepsilon_I^2+\varepsilon_Q^2)^{1/2}$ (dashed line) for the circular drive path (top row) and the straight path (bottom row) yielding the same dynamical phase.
}
\label{fig2}
\end{figure}

By applying a~$\pi/2$ pulse to the qubit \cite{Wallraff2005}, we initially prepare the system in the equal superposition of  $|0\rangle \otimes |g\rangle$ and $|0\rangle \otimes |e\rangle$, where $|0\rangle$ is the resonator vacuum state.
The adiabatic cycle takes the system into the state
\[
  \frac{1}{\sqrt{2}}(\exp(i\gamma^{(g)})|0\rangle \otimes |g\rangle + \exp(i\gamma^{(e)}) |0\rangle \otimes |e\rangle).
\] 
Note that due to the cyclicity and adiabaticity of the process, the resonator returns back to the vacuum state regardless of the state of the qubit and the two systems are again disentangled. We then finish the manipulation (Ramsey sequence; Fig.~\ref{fig2}(a)) with a~second $\pi/2$ pulse applied to the qubit, either in phase with the first one or shifted by $\pi/2$, and use dispersive readout \cite{bianchetti2009} to measure its excited state population. In this way we obtain the $x$ and $y$ projections of the qubit state Bloch vector $\langle\vec{\sigma}\rangle$, allowing us to calculate the phase difference $\gamma = \gamma^{(e)}-\gamma^{(g)}$ as the angle of rotation of the Bloch vector about the $z$ axis. This phase difference again contains a~dynamical part, related to the ac~Stark shift, and a~geometric contribution $\gamma_g = -2\Delta A$, where $\Delta A$ is the area enclosed between the coherent state paths for the qubit in the ground and in the excited state [Fig.~\ref{fig1}(c)]. Drive amplitudes typically used in our experiment result in up to $n\approx 20$ resonator photons, corresponding to $\gamma_g$ on the order of $2\pi$.

We control the resonator state to trace out different path shapes [Fig.~\ref{fig2}(b)] in clockwise and counterclockwise direction and measure the accumulated phase. For each of these paths we also measure the phase when leaving the amplitude modulation unchanged, but keeping the phase of the drive constant [Fig.~\ref{fig2}(c)]. The resulting straight paths yield the same dynamical phase as each of the original paths but no geometric phase since the area enclosed by them vanishes.

\begin{figure}
\includegraphics{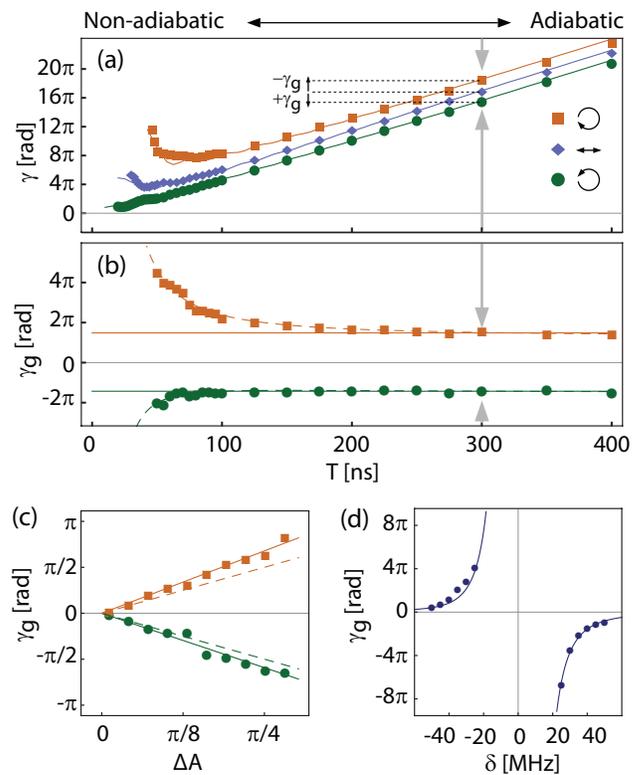}
\caption{
{(a)}
Measured total phase difference $\gamma$ for a~resonator pulse of fixed maximum amplitude $\varepsilon_0/2\pi\approx 370\ \mathrm{MHz}$, detuning $\delta/2\pi = 40\ \mathrm{MHz}$ and varying duration $T$, tracing a~counterclockwise circular (green circles), clockwise circular (orange squares) and straight trajectory (blue diamonds) with the same time-dependence of the drive amplitude. Solid lines show theory.
{(b)}
Dependence of the geometric phase on the drive pulse duration for the counterclockwise and clockwise trajectory. Solid lines show the adiabatic limit, dashed lines are fits with corrections proportional to $1/T$ and $1/T^2$. 
{(c)}
Geometric phase for a~counterclockwise and clockwise circular pulse with fixed duration $T = 300\ \mathrm{ns}$, detuning $\delta/2\pi = 40\ \mathrm{MHz}$ and varying maximum amplitude, plotted as a~function of the area $\Delta A$ enclosed between the coherent state trajectories, shown together with fitted linear functions (solid lines). Dashed lines represent the theoretical dependence $\gamma_g = -2 \Delta A$.
{(d)}
Measured geometric phase in the adiabatic limit as a~function of the detuning $\delta$, compared with the theoretical dependence (solid line).
}
\label{fig3}
\end{figure}

In this way, the phases are measured for different lengths $T$ of the drive pulse. In the adiabatic limit ($T\gtrsim 100\ \mathrm{ns}$), the dynamical phase, measured for the straight path, scales linearly with $T$ [Fig.~\ref{fig3}(a)], as expected. The geometric phase, evaluated as the difference between the phase for the 'area-enclosing' path and the dynamical phase, approaches a~constant value whose sign depends on the path orientation [Fig.~\ref{fig3}(b)]. As the cycle is traversed faster, and $T$ becomes comparable to $2\pi/|\delta| = 25\ \mathrm{ns}$, non-adiabatic effects become apparent. In the weakly non-adiabatic regime ($T\gtrsim 25\ \mathrm{ns}$), these are well described by an expansion in powers of $1/T$ \cite{berry1987}. Notably, the observed non-adiabatic corrections remain small even if the adiabaticity condition $T\gg 2\pi/|\delta|$ is clearly violated.

The scaling of $\gamma_g$ with the area $\Delta A$ (determined by the amplitude of the drive pulse) is observed to be linear [Fig.~\ref{fig3}(c)], in agreement with theory. We also verify this scaling by comparing the adiabatic geometric phase, obtained by extrapolating the measured geometric phases to $T\to\infty$, for different path shapes indicated in Fig.~\ref{fig2}(b). The ratios of the extracted phases to that of the circular path are $0.493\pm 0.016$ ($1/2$) for the semicircle, $0.647\pm 0.016$ ($2/\pi\approx 0.637$) for the square and $0.00\pm 0.07$ ($0$) for the figure-eight shape, in excellent agreement with the theoretical values stated in parentheses.

We have also measured the dependence of the geometric phase on the drive detuning $\delta$, observing its increase with decreasing $\delta$. This trend is explained by the larger field amplitudes and, thus, an increased geometric phase when driving the resonator closer to its resonance frequency.
Changing the sign of the detuning $\delta$ reverses the sign of the resulting phase since the relative size of the amplitudes of the resonator field for the two qubit states is reversed. The measured dependence  agrees well with these theoretical predictions [Fig.~\ref{fig3}(d)]. 

\begin{figure}
\includegraphics[scale=0.95]{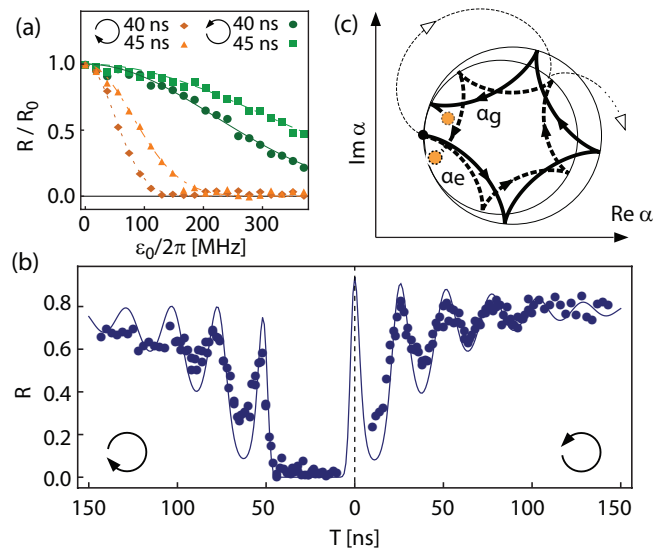}
\caption{
{(a)}
Relative length $R$ of the $xy$ Bloch vector projection versus the maximum amplitude $\varepsilon_0$ of the circular drive pulse for different lengths and orientations of the pulse, normalized to the projected Bloch vector length $R_0$ for $\varepsilon_0=0$. Solid lines show fitted Gaussian functions.
{(b)}
Length $R$ of the $xy$ Bloch vector projection as a~function of the resonator pulse length for a~fixed pulse amplitude $\varepsilon_0/2\pi\approx 190\ \mathrm{MHz}$ at $\delta=40\ \mathrm{MHz}$, corresponding to $n\approx 5$. The solid line is obtained from the analytically solvable evolution equation for the oscillator.
{(c)}
Trajectories of the resonator coherent state driven by a~circular pulse for the ground (solid thick line) and excited (dashed thick line) qubit state in the non-adiabatic regime. The different final quadratures $\alpha_g$ and $\alpha_e$ are indicated by yellow circles at the end of the paths. A~part of the trajectory for the opposite orientation of the drive pulse (dashed thin line), as well as the circular paths expected in the adiabatic limit (solid thin lines) are shown for comparison.
}
\label{fig4}
\end{figure}

Apart from the rotation of the qubit state Bloch vector about the $z$ axis, representing the accumulated phase difference, we also observe a~decrease in the length $R$ of its $xy$ projection, i.e.~dephasing of the qubit. The measured value of $R$ shows a~strong dependence both on the drive pulse amplitude [Fig.~\ref{fig4}(a)] and duration [Fig.~\ref{fig4}(b)]. The dephasing effect can be explained as a~result of non-adiabaticity which leads to entanglement between the oscillator and the qubit. For a~non-adiabatic drive cycle, the resonator coherent state does not follow the changes of the drive parameters [Fig.~\ref{fig4}(c)] and its trajectory exhibits periodic excursions from the adiabatic path, i.e., ringing. The system does not necessarily return to a~product state of the qubit and the resonator and in general, the final quadratures of the resonator field for the two qubit states differ. Therefore, the two subsystems remain entangled \cite{utami2008} and the reduced density matrix of the qubit corresponds to a~mixed state with a~Bloch vector length decreased by a~factor given by the overlap 
\begin{equation}
  |\langle\alpha_g|\alpha_e\rangle| = \exp(-|\alpha_g-\alpha_e|^2/2)
\end{equation}
of the two final resonator states. Alternatively, this dephasing effect can be understood as measurement-induced dephasing -- a~result of the resonator field extracting information about the qubit state \cite{gambetta2006}.

The Gaussian dependence of $R$ on the drive amplitude [Fig.~\ref{fig4}(a)] follows directly from Eq.~(1) and the proportionality between $\alpha$ and $\varepsilon$. The fall-off of $R$ is faster for shorter, less adiabatic pulses as they result in larger separation between the final field quadratures and hence stronger dephasing. Interestingly, the magnitude of the dephasing effect shows oscillations in the evolution time $T$ and also depends on the orientation of the drive cycle, as seen in Fig.~\ref{fig4}(b). The asymmetric behavior has a~simple explanation in terms of the Fourier transform of the drive signal. For the circular path, the two dominant Fourier components have frequencies $\omega$ and $\omega \pm 2\pi/T$, where the signs $+$ and $-$ correspond to the clockwise and counterclockwise orientation of the path, respectively.
In one of these two cases, the resonator is driven closer to its resonance frequency, resulting in stronger dephasing. For fast clockwise drive cycles ($T\lesssim 50\ \mathrm{ns}$) the qubit is almost completely dephased and the projected Bloch vector length $R$ approaches zero.  The oscillatory behaviour of $R$ with a period of $25\ \mathrm{ns}$ corresponding to $\delta/2\pi = 40\ \mathrm{MHz}$, is due to the ringing of the resonator field. For cycle periods corresponding to the maxima of the Bloch vector length $R$, at which the ringing frequencies are commensurate, the resonator state path is cyclic and dephasing is minimized. These particular periods could be used to realize fast non-adiabatic geometric gates.

The high level of control over the coherent state of the resonator field has allowed us to measure its geometric phase for a~wide range of path shapes in the adiabatic and the non-adiabatic regime. The characteristic features of the geometric phase that we have observed -- independence on dynamical quantities and scaling with the area enclosed by the trajectory in the parameter space -- are in good agreement with theory. Non-adiabatic effects introduce corrections to the geometric phase and dephasing of the reduced qubit state due to residual qubit--resonator entanglement. In the future, our system can serve as a~valuable tool for investigations of non-adiabatic geometric phases to shed light on their robustness and potential use in fast geometric quantum gates.

\emph{Acknowledgements:} We thank M. M\"ott\"onen, R. Fazio, A. Shnirman and A. Clerk for useful comments on the manuscript. Supported by the EU project GEOMDISS, the Austrian Science Foundation (S.~F.) and the Swiss National Science Foundation (SNSF).


\begin{thebibliography}{36}%
\makeatletter
\providecommand \@ifxundefined [1]{%
 \@ifx{#1\undefined}
}%
\providecommand \@ifnum [1]{%
 \ifnum #1\expandafter \@firstoftwo
 \else \expandafter \@secondoftwo
 \fi
}%
\providecommand \@ifx [1]{%
 \ifx #1\expandafter \@firstoftwo
 \else \expandafter \@secondoftwo
 \fi
}%
\providecommand \natexlab [1]{#1}%
\providecommand \enquote  [1]{``#1''}%
\providecommand \bibnamefont  [1]{#1}%
\providecommand \bibfnamefont [1]{#1}%
\providecommand \citenamefont [1]{#1}%
\providecommand \href@noop [0]{\@secondoftwo}%
\providecommand \href [0]{\begingroup \@sanitize@url \@href}%
\providecommand \@href[1]{\@@startlink{#1}\@@href}%
\providecommand \@@href[1]{\endgroup#1\@@endlink}%
\providecommand \@sanitize@url [0]{\catcode `\\12\catcode `\$12\catcode
  `\&12\catcode `\#12\catcode `\^12\catcode `\_12\catcode `\%12\relax}%
\providecommand \@@startlink[1]{}%
\providecommand \@@endlink[0]{}%
\providecommand \url  [0]{\begingroup\@sanitize@url \@url }%
\providecommand \@url [1]{\endgroup\@href {#1}{\urlprefix }}%
\providecommand \urlprefix  [0]{URL }%
\providecommand \Eprint [0]{\href }%
\providecommand \doibase [0]{http://dx.doi.org/}%
\providecommand \selectlanguage [0]{\@gobble}%
\providecommand \bibinfo  [0]{\@secondoftwo}%
\providecommand \bibfield  [0]{\@secondoftwo}%
\providecommand \translation [1]{[#1]}%
\providecommand \BibitemOpen [0]{}%
\providecommand \bibitemStop [0]{}%
\providecommand \bibitemNoStop [0]{.\EOS\space}%
\providecommand \EOS [0]{\spacefactor3000\relax}%
\providecommand \BibitemShut  [1]{\csname bibitem#1\endcsname}%
\let\auto@bib@innerbib\@empty
\bibitem [{\citenamefont {Messiah}(1962)}]{MessiahII}%
  \BibitemOpen
  \bibfield  {author} {\bibinfo {author} {\bibfnamefont {A.}~\bibnamefont
  {Messiah}},\ }\href@noop {} {\emph {\bibinfo {title} {Quantum Mechanics,
  Volume II}}}\ (\bibinfo  {publisher} {North-Holland Publishing Company,
  Amsterdam},\ \bibinfo {year} {1962})\BibitemShut {NoStop}%
\bibitem [{\citenamefont {Berry}(1984)}]{berry1984}%
  \BibitemOpen
  \bibfield  {author} {\bibinfo {author} {\bibfnamefont {M.}~\bibnamefont
  {Berry}},\ }\href@noop {} {\bibfield  {journal} {\bibinfo  {journal} {Proc.
  R. Soc. Lond. A}\ }\textbf {\bibinfo {volume} {392}},\ \bibinfo {pages} {45–}
  (\bibinfo {year} {1984})}\BibitemShut {NoStop}%
\bibitem [{\citenamefont {Bitter}\ and\ \citenamefont
  {Dubbers}(1987)}]{bitter1987}%
  \BibitemOpen
  \bibfield  {author} {\bibinfo {author} {\bibfnamefont {T.}~\bibnamefont
  {Bitter}}\ and\ \bibinfo {author} {\bibfnamefont {D.}~\bibnamefont
  {Dubbers}},\ }\href {\doibase 10.1103/PhysRevLett.59.251} {\bibfield
  {journal} {\bibinfo  {journal} {Phys. Rev. Lett.}\ }\textbf {\bibinfo
  {volume} {59}},\ \bibinfo {pages} {251} (\bibinfo {year} {1987})}\BibitemShut
  {NoStop}%
\bibitem [{\citenamefont {Suter}\ \emph {et~al.}(1988)\citenamefont {Suter},
  \citenamefont {Mueller},\ and\ \citenamefont {Pines}}]{suter1988}%
  \BibitemOpen
  \bibfield  {author} {\bibinfo {author} {\bibfnamefont {D.}~\bibnamefont
  {Suter}}, \bibinfo {author} {\bibfnamefont {K.~T.}\ \bibnamefont {Mueller}},
  \ and\ \bibinfo {author} {\bibfnamefont {A.}~\bibnamefont {Pines}},\ }\href
  {\doibase 10.1103/PhysRevLett.60.1218} {\bibfield  {journal} {\bibinfo
  {journal} {Phys. Rev. Lett.}\ }\textbf {\bibinfo {volume} {60}},\ \bibinfo
  {pages} {1218} (\bibinfo {year} {1988})}\BibitemShut {NoStop}%
\bibitem [{\citenamefont {Jones}\ \emph {et~al.}(2000)\citenamefont {Jones},
  \citenamefont {Vedral}, \citenamefont {Ekert},\ and\ \citenamefont
  {Castagnoli}}]{jones2000}%
  \BibitemOpen
  \bibfield  {author} {\bibinfo {author} {\bibfnamefont {J.}~\bibnamefont
  {Jones}}, \bibinfo {author} {\bibfnamefont {V.}~\bibnamefont {Vedral}},
  \bibinfo {author} {\bibfnamefont {A.}~\bibnamefont {Ekert}}, \ and\ \bibinfo
  {author} {\bibfnamefont {G.}~\bibnamefont {Castagnoli}},\ }\href@noop {}
  {\bibfield  {journal} {\bibinfo  {journal} {Nature}\ }\textbf {\bibinfo
  {volume} {403}},\ \bibinfo {pages} {869} (\bibinfo {year}
  {2000})}\BibitemShut {NoStop}%
\bibitem [{\citenamefont {Leek}\ \emph {et~al.}(2007)\citenamefont {Leek},
  \citenamefont {Fink}, \citenamefont {Blais}, \citenamefont {Bianchetti},
  \citenamefont {G\"{o}ppl}, \citenamefont {Gambetta}, \citenamefont
  {Schuster}, \citenamefont {Frunzio}, \citenamefont {Schoelkopf},\ and\
  \citenamefont {A.Wallraff}}]{leek2007}%
  \BibitemOpen
  \bibfield  {author} {\bibinfo {author} {\bibfnamefont {P.}~\bibnamefont
  {Leek}}, \emph{et~al.},\ }\href@noop {} {\bibfield  {journal} {\bibinfo  {journal}
  {Science}\ }\textbf {\bibinfo {volume} {318}},\ \bibinfo {pages} {1889}
  (\bibinfo {year} {2007})}\BibitemShut {NoStop}%
\bibitem [{\citenamefont {Neeley}\ \emph {et~al.}(2009)\citenamefont {Neeley},
  \citenamefont {Ansmann}, \citenamefont {Bialczak}, \citenamefont {Hofheinz},
  \citenamefont {Lucero}, \citenamefont {O'Connell}, \citenamefont {Sank},
  \citenamefont {Wang}, \citenamefont {Wenner}, \citenamefont {Cleland},
  \citenamefont {Geller},\ and\ \citenamefont {Martinis}}]{Neeley2009}%
  \BibitemOpen
  \bibfield  {author} {\bibinfo {author} {\bibfnamefont {M.}~\bibnamefont
  {Neeley}}, \emph{et~al.},\ }\href@noop {} {\bibfield
  {journal} {\bibinfo  {journal} {Science}\ }\textbf {\bibinfo {volume}
  {325}},\ \bibinfo {pages} {722} (\bibinfo {year} {2009})}\BibitemShut
  {NoStop}%
\bibitem [{\citenamefont {Leibfried}\ \emph {et~al.}(2003)\citenamefont
  {Leibfried}, \citenamefont {DeMarco}, \citenamefont {Meyer}, \citenamefont
  {Lucas}, \citenamefont {Barrett}, \citenamefont {Britton}, \citenamefont
  {Itano}, \citenamefont {Jelenkovic}, \citenamefont {Langer}, \citenamefont
  {Rosenband},\ and\ \citenamefont {Wineland}}]{leibfried2003}%
  \BibitemOpen
  \bibfield  {author} {\bibinfo {author} {\bibfnamefont {D.}~\bibnamefont
  {Leibfried}}, \emph{et~al.},\ }\href@noop {} {\bibfield  {journal}
  {\bibinfo  {journal} {Nature}\ }\textbf {\bibinfo {volume} {422}},\ \bibinfo
  {pages} {412} (\bibinfo {year} {2003})}\BibitemShut {NoStop}%
\bibitem [{\citenamefont {{De Chiara}}\ and\ \citenamefont
  {Palma}(2003)}]{dechiara2003}%
  \BibitemOpen
  \bibfield  {author} {\bibinfo {author} {\bibfnamefont {G.}~\bibnamefont {{De
  Chiara}}}\ and\ \bibinfo {author} {\bibfnamefont {G.}~\bibnamefont {Palma}},\
  }\href@noop {} {\bibfield  {journal} {\bibinfo  {journal} {Phys. Rev. Lett.}\
  }\textbf {\bibinfo {volume} {91}},\ \bibinfo {pages} {090404} (\bibinfo
  {year} {2003})}\BibitemShut {NoStop}%
\bibitem [{\citenamefont {Carollo}\ \emph {et~al.}(2004)\citenamefont
  {Carollo}, \citenamefont {Fuentes-Guridi}, \citenamefont {Santos},\ and\
  \citenamefont {Vedral}}]{carollo2004}%
  \BibitemOpen
  \bibfield  {author} {\bibinfo {author} {\bibfnamefont {A.}~\bibnamefont
  {Carollo}}, \bibinfo {author} {\bibfnamefont {I.}~\bibnamefont
  {Fuentes-Guridi}}, \bibinfo {author} {\bibfnamefont {M.~F.}\ \bibnamefont
  {Santos}}, \ and\ \bibinfo {author} {\bibfnamefont {V.}~\bibnamefont
  {Vedral}},\ }\href@noop {} {\bibfield  {journal} {\bibinfo  {journal} {Phys.
  Rev. Lett.}\ }\textbf {\bibinfo {volume} {92}},\ \bibinfo {pages} {020402}
  (\bibinfo {year} {2004})}\BibitemShut {NoStop}%
\bibitem [{\citenamefont {Whitney}\ \emph {et~al.}(2005)\citenamefont
  {Whitney}, \citenamefont {Makhlin}, \citenamefont {Shnirman},\ and\
  \citenamefont {Gefen}}]{whitney2005}%
  \BibitemOpen
  \bibfield  {author} {\bibinfo {author} {\bibfnamefont {R.}~\bibnamefont
  {Whitney}}, \bibinfo {author} {\bibfnamefont {Y.}~\bibnamefont {Makhlin}},
  \bibinfo {author} {\bibfnamefont {A.}~\bibnamefont {Shnirman}}, \ and\
  \bibinfo {author} {\bibfnamefont {Y.}~\bibnamefont {Gefen}},\ }\href@noop {}
  {\bibfield  {journal} {\bibinfo  {journal} {Phys. Rev. Lett.}\ }\textbf
  {\bibinfo {volume} {94}},\ \bibinfo {pages} {070407} (\bibinfo {year}
  {2005})}\BibitemShut {NoStop}%
\bibitem [{\citenamefont {Blais}\ and\ \citenamefont
  {Tremblay}(2003)}]{blais2003}%
  \BibitemOpen
  \bibfield  {author} {\bibinfo {author} {\bibfnamefont {A.}~\bibnamefont
  {Blais}}\ and\ \bibinfo {author} {\bibfnamefont {A.-M.}\ \bibnamefont
  {Tremblay}},\ }\href@noop {} {\bibfield  {journal} {\bibinfo  {journal}
  {Phys. Rev. A}\ }\textbf {\bibinfo {volume} {67}},\ \bibinfo {pages} {012308}
  (\bibinfo {year} {2003})}\BibitemShut {NoStop}%
\bibitem [{\citenamefont {Thomas}\ \emph {et~al.}(2011)\citenamefont {Thomas},
  \citenamefont {Lababidi},\ and\ \citenamefont {Tian}}]{thomas2011}%
  \BibitemOpen
  \bibfield  {author} {\bibinfo {author} {\bibfnamefont {J.~T.}\ \bibnamefont
  {Thomas}}, \bibinfo {author} {\bibfnamefont {M.}~\bibnamefont {Lababidi}}, \
  and\ \bibinfo {author} {\bibfnamefont {M.}~\bibnamefont {Tian}},\ }\href@noop
  {} {\bibfield  {journal} {\bibinfo  {journal} {Phys. Rev. A}\ }\textbf
  {\bibinfo {volume} {84}},\ \bibinfo {pages} {042335} (\bibinfo {year}
  {2011})}\BibitemShut {NoStop}%
\bibitem [{\citenamefont {Fuentes-Guridi}\ \emph {et~al.}(2000)\citenamefont
  {Fuentes-Guridi}, \citenamefont {Bose},\ and\ \citenamefont
  {Vedral}}]{fuentesguridi2000}%
  \BibitemOpen
  \bibfield  {author} {\bibinfo {author} {\bibfnamefont {I.}~\bibnamefont
  {Fuentes-Guridi}}, \bibinfo {author} {\bibfnamefont {S.}~\bibnamefont
  {Bose}}, \ and\ \bibinfo {author} {\bibfnamefont {V.}~\bibnamefont
  {Vedral}},\ }\href@noop {} {\bibfield  {journal} {\bibinfo  {journal} {Phys.
  Rev. Lett.}\ }\textbf {\bibinfo {volume} {85}},\ \bibinfo {pages} {5018}
  (\bibinfo {year} {2000})}\BibitemShut {NoStop}%
\bibitem [{\citenamefont {Wang}\ and\ \citenamefont {Keiji}(2001)}]{wang2001A}%
  \BibitemOpen
  \bibfield  {author} {\bibinfo {author} {\bibfnamefont {X.}~\bibnamefont
  {Wang}}\ and\ \bibinfo {author} {\bibfnamefont {M.}~\bibnamefont {Keiji}},\
  }\href@noop {} {\bibfield  {journal} {\bibinfo  {journal} {Phys. Rev. Lett.}\
  }\textbf {\bibinfo {volume} {87}},\ \bibinfo {pages} {097901} (\bibinfo
  {year} {2001})}\BibitemShut {NoStop}%
\bibitem [{\citenamefont {Zheng}(2004)}]{zheng2004}%
  \BibitemOpen
  \bibfield  {author} {\bibinfo {author} {\bibfnamefont {S.-B.}\ \bibnamefont
  {Zheng}},\ }\href@noop {} {\bibfield  {journal} {\bibinfo  {journal} {Phys.
  Rev. A}\ }\textbf {\bibinfo {volume} {70}},\ \bibinfo {pages} {052320}
  (\bibinfo {year} {2004})}\BibitemShut {NoStop}%
\bibitem [{\citenamefont {Garc\'{i}a-Ripoll}\ \emph {et~al.}(2005)\citenamefont
  {Garc\'{i}a-Ripoll}, \citenamefont {Zoller},\ and\ \citenamefont
  {Cirac}}]{garcia-ripoll2005}%
  \BibitemOpen
  \bibfield  {author} {\bibinfo {author} {\bibfnamefont {J.}~\bibnamefont
  {Garc\'{i}a-Ripoll}}, \bibinfo {author} {\bibfnamefont {P.}~\bibnamefont
  {Zoller}}, \ and\ \bibinfo {author} {\bibfnamefont {J.~I.}\ \bibnamefont
  {Cirac}},\ }\href@noop {} {\bibfield  {journal} {\bibinfo  {journal} {Phys.
  Rev. A}\ }\textbf {\bibinfo {volume} {71}},\ \bibinfo {pages} {062309}
  (\bibinfo {year} {2005})}\BibitemShut {NoStop}%
\bibitem [{\citenamefont {Blais}\ \emph {et~al.}(2007)\citenamefont {Blais},
  \citenamefont {Gambetta}, \citenamefont {Wallraff}, \citenamefont {Schuster},
  \citenamefont {Girvin}, \citenamefont {Devoret},\ and\ \citenamefont
  {Schoelkopf}}]{blais2007}%
  \BibitemOpen
  \bibfield  {author} {\bibinfo {author} {\bibfnamefont {A.}~\bibnamefont
  {Blais}}, \emph{et~al.},\ }\href@noop {} {\bibfield  {journal}
  {\bibinfo  {journal} {Phys. Rev. A}\ }\textbf {\bibinfo {volume} {75}},\
  \bibinfo {pages} {032329} (\bibinfo {year} {2007})}\BibitemShut {NoStop}%
\bibitem [{\citenamefont {Zanardi}\ and\ \citenamefont
  {Rasetti}(1999)}]{zanardi1999}%
  \BibitemOpen
  \bibfield  {author} {\bibinfo {author} {\bibfnamefont {P.}~\bibnamefont
  {Zanardi}}\ and\ \bibinfo {author} {\bibfnamefont {M.}~\bibnamefont
  {Rasetti}},\ }\href@noop {} {\bibfield  {journal} {\bibinfo  {journal} {Phys.
  Lett. A}\ }\textbf {\bibinfo {volume} {264}},\ \bibinfo {pages} {94}
  (\bibinfo {year} {1999})}\BibitemShut {NoStop}%
\bibitem [{\citenamefont {Duan}\ \emph {et~al.}(2001)\citenamefont {Duan},
  \citenamefont {Cirac},\ and\ \citenamefont {Zoller}}]{duan2001}%
  \BibitemOpen
  \bibfield  {author} {\bibinfo {author} {\bibfnamefont {L.-M.}\ \bibnamefont
  {Duan}}, \bibinfo {author} {\bibfnamefont {J.~I.}\ \bibnamefont {Cirac}}, \
  and\ \bibinfo {author} {\bibfnamefont {P.}~\bibnamefont {Zoller}},\
  }\href@noop {} {\bibfield  {journal} {\bibinfo  {journal} {Science}\ }\textbf
  {\bibinfo {volume} {292}},\ \bibinfo {pages} {1695} (\bibinfo {year}
  {2001})}\BibitemShut {NoStop}%
\bibitem [{\citenamefont {Pachos}(2002)}]{pachos2002}%
  \BibitemOpen
  \bibfield  {author} {\bibinfo {author} {\bibfnamefont {J.}~\bibnamefont
  {Pachos}},\ }\href@noop {} {\bibfield  {journal} {\bibinfo  {journal} {Phys.
  Rev. A}\ }\textbf {\bibinfo {volume} {66}},\ \bibinfo {pages} {042318}
  (\bibinfo {year} {2002})}\BibitemShut {NoStop}%
\bibitem [{\citenamefont {Kamleitner}\ \emph {et~al.}(2011)\citenamefont
  {Kamleitner}, \citenamefont {Solinas}, \citenamefont {M\"{u}ller},
  \citenamefont {Shnirman},\ and\ \citenamefont
  {M\"{o}tt\"{o}nen}}]{kamleitner2011}%
  \BibitemOpen
  \bibfield  {author} {\bibinfo {author} {\bibfnamefont {I.}~\bibnamefont
  {Kamleitner}}, \emph{et~al.},\ }\href@noop
  {} {\bibfield  {journal} {\bibinfo  {journal} {Phys. Rev. B}\ }\textbf
  {\bibinfo {volume} {83}},\ \bibinfo {pages} {214518} (\bibinfo {year}
  {2011})}\BibitemShut {NoStop}%
\bibitem [{\citenamefont {Filipp}\ \emph {et~al.}(2009)\citenamefont {Filipp},
  \citenamefont {Klepp}, \citenamefont {Hasegawa}, \citenamefont
  {Plonka-Spehr}, \citenamefont {Schmidt}, \citenamefont {Geltenbort},\ and\
  \citenamefont {Rauch}}]{filipp2009G}%
  \BibitemOpen
  \bibfield  {author} {\bibinfo {author} {\bibfnamefont {S.}~\bibnamefont
  {Filipp}}, \emph{et~al.},\ }\href@noop {} {\bibfield  {journal} {\bibinfo
  {journal} {Phys. Rev. Lett.}\ }\textbf {\bibinfo {volume} {102}},\ \bibinfo
  {pages} {030404} (\bibinfo {year} {2009})}\BibitemShut {NoStop}%
\bibitem [{\citenamefont {Cucchietti}\ \emph {et~al.}(2010)\citenamefont
  {Cucchietti}, \citenamefont {Zhang}, \citenamefont {Lombardo}, \citenamefont
  {Villar},\ and\ \citenamefont {Laflamme}}]{cucchietti2010}%
  \BibitemOpen
  \bibfield  {author} {\bibinfo {author} {\bibfnamefont {F.}~\bibnamefont
  {Cucchietti}}, \emph{et~al.},\ }\href@noop {}
  {\bibfield  {journal} {\bibinfo  {journal} {Phys. Rev. Lett.}\ }\textbf
  {\bibinfo {volume} {105}},\ \bibinfo {pages} {240406} (\bibinfo {year}
  {2010})}\BibitemShut {NoStop}%
\bibitem [{\citenamefont {Blais}\ \emph {et~al.}(2004)\citenamefont {Blais},
  \citenamefont {Huang}, \citenamefont {Wallraff}, \citenamefont {Girvin},\
  and\ \citenamefont {Schoelkopf}}]{blais2004}%
  \BibitemOpen
  \bibfield  {author} {\bibinfo {author} {\bibfnamefont {A.}~\bibnamefont
  {Blais}}, \emph{et~al.},\ }\href@noop {} {\bibfield
  {journal} {\bibinfo  {journal} {Phys. Rev. A}\ }\textbf {\bibinfo {volume}
  {69}},\ \bibinfo {pages} {062320} (\bibinfo {year} {2004})}\BibitemShut
  {NoStop}%
\bibitem [{\citenamefont {Vacanti}\ \emph {et~al.}(2011)\citenamefont
  {Vacanti}, \citenamefont {Fazio}, \citenamefont {Kim}, \citenamefont {Palma},
  \citenamefont {Paternostro},\ and\ \citenamefont {Vedral}}]{Vacanti2011}%
  \BibitemOpen
  \bibfield  {author} {\bibinfo {author} {\bibfnamefont {G.}~\bibnamefont
  {Vacanti}}, \emph{et~al.},\ }\href@noop {} {\bibfield  {journal} {\bibinfo
  {journal} {arXiv:1108.0701v1 [quant-ph]}\ } (\bibinfo {year}
  {2011})}\BibitemShut {NoStop}%
\bibitem [{\citenamefont {Aharonov}\ and\ \citenamefont
  {Anandan}(1987)}]{Aharonov1987}%
  \BibitemOpen
  \bibfield  {author} {\bibinfo {author} {\bibfnamefont {Y.}~\bibnamefont
  {Aharonov}}\ and\ \bibinfo {author} {\bibfnamefont {J.}~\bibnamefont
  {Anandan}},\ }\href@noop {} {\bibfield  {journal} {\bibinfo  {journal} {Phys.
  Rev. Lett.}\ }\textbf {\bibinfo {volume} {58}},\ \bibinfo {pages} {1593–}
  (\bibinfo {year} {1987})}\BibitemShut {NoStop}%
\bibitem [{\citenamefont {Zhu}\ and\ \citenamefont {Wang}(2002)}]{zhuPRL2002}%
  \BibitemOpen
  \bibfield  {author} {\bibinfo {author} {\bibfnamefont {S.}~\bibnamefont
  {Zhu}}\ and\ \bibinfo {author} {\bibfnamefont {Z.}~\bibnamefont {Wang}},\
  }\href@noop {} {\bibfield  {journal} {\bibinfo  {journal} {Phys. Rev. Lett.}\
  }\textbf {\bibinfo {volume} {89}},\ \bibinfo {pages} {097902} (\bibinfo
  {year} {2002})}\BibitemShut {NoStop}%
\bibitem [{\citenamefont {Koch}\ \emph {et~al.}(2007)\citenamefont {Koch},
  \citenamefont {Yu}, \citenamefont {Gambetta}, \citenamefont {Houck},
  \citenamefont {Schuster}, \citenamefont {Majer}, \citenamefont {Blais},
  \citenamefont {Devoret}, \citenamefont {Girvin},\ and\ \citenamefont
  {Schoelkopf}}]{koch2007}%
  \BibitemOpen
  \bibfield  {author} {\bibinfo {author} {\bibfnamefont {J.}~\bibnamefont
  {Koch}}, \emph{et~al.},\ }\href@noop {} {\bibfield  {journal}
  {\bibinfo  {journal} {Phys. Rev. A}\ }\textbf {\bibinfo {volume} {76}},\
  \bibinfo {pages} {042319} (\bibinfo {year} {2007})}\BibitemShut {NoStop}%
\bibitem [{\citenamefont {Schuster}\ \emph {et~al.}(2007)\citenamefont
  {Schuster}, \citenamefont {Houck}, \citenamefont {Schreier}, \citenamefont
  {Wallraff}, \citenamefont {Gambetta}, \citenamefont {Blais}, \citenamefont
  {Frunzio}, \citenamefont {Majer}, \citenamefont {Johnson}, \citenamefont
  {Devoret}, \citenamefont {Girvin},\ and\ \citenamefont
  {Schoelkopf}}]{schuster2007}%
  \BibitemOpen
  \bibfield  {author} {\bibinfo {author} {\bibfnamefont {D.}~\bibnamefont
  {Schuster}}, \emph{et~al.},\ }\href@noop {} {\bibfield  {journal}
  {\bibinfo  {journal} {Nature}\ }\textbf {\bibinfo {volume} {445}},\ \bibinfo
  {pages} {515} (\bibinfo {year} {2007})}\BibitemShut {NoStop}%
\bibitem [{\citenamefont {Chaturvedi}\ \emph {et~al.}(1987)\citenamefont
  {Chaturvedi}, \citenamefont {Sriram},\ and\ \citenamefont
  {Srinivasan}}]{chaturvedi1987}%
  \BibitemOpen
  \bibfield  {author} {\bibinfo {author} {\bibfnamefont {S.}~\bibnamefont
  {Chaturvedi}}, \bibinfo {author} {\bibfnamefont {M.}~\bibnamefont {Sriram}},
  \ and\ \bibinfo {author} {\bibfnamefont {V.}~\bibnamefont {Srinivasan}},\
  }\href@noop {} {\bibfield  {journal} {\bibinfo  {journal} {J. Phys. A: Math.
  Gen.}\ }\textbf {\bibinfo {volume} {20}},\ \bibinfo {pages} {L1071} (\bibinfo
  {year} {1987})}\BibitemShut {NoStop}%
\bibitem [{\citenamefont {Wallraff}\ \emph {et~al.}(2005)\citenamefont
  {Wallraff}, \citenamefont {Schuster}, \citenamefont {Blais}, \citenamefont
  {Frunzio}, \citenamefont {J.~Majer}, \citenamefont {Girvin},\ and\
  \citenamefont {Schoelkopf}}]{Wallraff2005}%
  \BibitemOpen
  \bibfield  {author} {\bibinfo {author} {\bibfnamefont {A.}~\bibnamefont
  {Wallraff}}, \emph{et~al.},\ }\href@noop {} {\bibfield
  {journal} {\bibinfo  {journal} {Phys. Rev. Lett.}\ }\textbf {\bibinfo
  {volume} {95}},\ \bibinfo {pages} {060501} (\bibinfo {year}
  {2005})}\BibitemShut {NoStop}%
\bibitem [{\citenamefont {Bianchetti}\ \emph {et~al.}(2009)\citenamefont
  {Bianchetti}, \citenamefont {Filipp}, \citenamefont {Baur}, \citenamefont
  {Fink}, \citenamefont {G\"{o}ppl}, \citenamefont {Leek}, \citenamefont
  {Steffen}, \citenamefont {Blais},\ and\ \citenamefont
  {Wallraff}}]{bianchetti2009}%
  \BibitemOpen
  \bibfield  {author} {\bibinfo {author} {\bibfnamefont {R.}~\bibnamefont
  {Bianchetti}}, \emph{et~al.},\
  }\href@noop {} {\bibfield  {journal} {\bibinfo  {journal} {Phys. Rev. A}\
  }\textbf {\bibinfo {volume} {80}},\ \bibinfo {pages} {043840} (\bibinfo
  {year} {2009})}\BibitemShut {NoStop}%
\bibitem [{\citenamefont {Berry}(1987)}]{berry1987}%
  \BibitemOpen
  \bibfield  {author} {\bibinfo {author} {\bibfnamefont {M.}~\bibnamefont
  {Berry}},\ }\href@noop {} {\bibfield  {journal} {\bibinfo  {journal} {Proc.
  R. Soc. Lond. A}\ }\textbf {\bibinfo {volume} {414}},\ \bibinfo {pages} {31–}
  (\bibinfo {year} {1987})}\BibitemShut {NoStop}%
\bibitem [{\citenamefont {Utami}\ and\ \citenamefont
  {Clerk}(2008)}]{utami2008}%
  \BibitemOpen
  \bibfield  {author} {\bibinfo {author} {\bibfnamefont {D.~W.}\ \bibnamefont
  {Utami}}\ and\ \bibinfo {author} {\bibfnamefont {A.}~\bibnamefont {Clerk}},\
  }\href@noop {} {\bibfield  {journal} {\bibinfo  {journal} {Phys. Rev. A}\
  }\textbf {\bibinfo {volume} {78}},\ \bibinfo {pages} {042323} (\bibinfo
  {year} {2008})}\BibitemShut {NoStop}%
\bibitem [{\citenamefont {Gambetta}\ \emph {et~al.}(2006)\citenamefont
  {Gambetta}, \citenamefont {Blais}, \citenamefont {Schuster}, \citenamefont
  {Wallraff}, \citenamefont {Frunzio}, \citenamefont {Majer}, \citenamefont
  {Devoret}, \citenamefont {Girvin},\ and\ \citenamefont
  {Schoelkopf}}]{gambetta2006}%
  \BibitemOpen
  \bibfield  {author} {\bibinfo {author} {\bibfnamefont {J.}~\bibnamefont
  {Gambetta}}, \emph{et~al.},\ }\href@noop {} {\bibfield  {journal}
  {\bibinfo  {journal} {Phys. Rev. A}\ }\textbf {\bibinfo {volume} {74}},\
  \bibinfo {pages} {042318} (\bibinfo {year} {2006})}\BibitemShut {NoStop}%
\end{thebibliography}
\end{document}